\documentstyle[preprint,aps]{revtex}
\begin{document}
\draft

\title{Critical Dynamics of the Contact Process with Quenched Disorder}
\author{Adriana G. Moreira$^{\dagger,a}$ and Ronald Dickman$^{\ddagger,b}$}
\address{$^{\dagger}$Departamento de F\'\i sica, UFMG, CP 702, CEP 30161-970, 
Belo Horizonte, Brazil \\ and \\
$^{\ddagger}$Department of Physics and Astronomy, Lehman College, CUNY,
Bronx, NY 10468-1589, USA}
\date{\today}
\maketitle
\begin{abstract}
We study critical spreading dynamics in the 
two-dimensional contact process (CP) with quenched disorder in the 
form of random dilution.  In the pure model, spreading from a single particle
at the critical point $\lambda_c$ is characterized by the critical exponents 
of directed percolation: in $2+1$ dimensions, $\delta = 0.46$, $\eta = 0.214$, 
and $z = 1.13$.  Disorder
causes a dramatic change in the critical exponents, to
$\delta \simeq 0.60$, $\eta \simeq -0.42$, and $z \simeq 0.24$.  
These exponents govern spreading following a long crossover period.  
The usual hyperscaling relation, $4 \delta + 2 \eta = d z$, is
violated. Our results
support the conjecture by Bramson, Durrett, and Schonmann 
[Ann. Prob. {\bf 19}, 960 (1991)],
that  in two or more dimensions the disordered CP 
has only a single phase transition. 
\vspace {0.3truecm}

\end{abstract}
\vspace{1.0truecm}
 
\newpage
 
Phase transitions between an absorbing state  
(one, that is, admitting no further evolution), 
and an active regime occur in models of  
autocatalytic chemical reactions, epidemics, 
and transport in disordered media.  Critical phenomena 
attending absorbing-state transitions show a high degree 
of universality, that has been characterized rather precisely 
in studies of the contact process (CP) and of directed percolation (DP)
\cite{GRASS,KINZEL,CP92}. 
Since many-particle systems often incorporate 
frozen-in randomness, it is natural to investigate its effect 
on an absorbing-state transition.
Some time ago, Noest studied the change in the static critical behavior of  
DP due to quenched disorder \cite{NOEST}.   
In this work we examine its effect on {\em time-dependent} critical 
phenomena at an absorbing-state transition.  We focus on the two-dimensional 
CP, a simple lattice model of an epidemic \cite{TEHARRIS}.  
Our prime interest is in 
the effect of disorder on the spread of the critical process from a seed. 

At a critical creation rate, $\lambda_c$,
the pure CP exhibits a second-order phase transition 
characterized by the same critical exponents  
as DP \cite{GRASS}. The well-known Harris criterion 
\cite{HARRIS74,BERKER} states that disorder changes the critical exponents
if  $d\nu_{\perp}\leq 2$,
where $d$ is the dimensionality and $\nu_{\perp}$ is the correlation-length
exponent of the pure model. Since $\nu_{\perp} \simeq 0.73$ 
for DP in 2+1 dimensions,
we expect quenched disorder to be relevant in the CP.
Indeed, Noest's simulations of  one- and two-dimensional
stochastic cellular automata (belonging to the DP class), yielded static 
critical exponents quite distinct from
those of DP, when the models were modified 
to incorporate quenched randomness \cite{NOEST}. A field-theoretic
study by Obukhov \cite{OBU} yielded qualitatively consistent results.  
Marques studied the effects of dilution on the CP and related models in a 
mean-field renormalization group study, obtaining a 
phase diagram in good agreement with simulation \cite{MARQUES}.

The effect of disorder on time-dependent critical behavior at an absorbing-state
transition has received little attention.  The only study we are aware of, by
Bramson, Durrett, and Schonmann, showed that the 
one-dimensional CP in a random environment has
a new phase in which the process survives but 
spreads more slowly than linearly with time, t \cite{BRAMSON}.  
(In the pure CP the radius of the active region grows $\propto t$
for any $\lambda > \lambda_c$.)  

In this paper we report on extensive simulations of time-dependent 
critical behavior in the two-dimensional DCP.  The model is defined as follows.
Each site of the square lattice ${\cal Z}^2$ 
is either vacant or occupied by a particle.
Particles are created at vacant sites at rate $\lambda n /4$, where $n$ is
the number of occupied nearest-neighbors, and
are annihilated at unit rate, independent of the surrounding
configuration. The order parameter is the particle
density $\rho$; it vanishes in the vacuum state, which is absorbing.
As $\lambda$ is increased beyond $\lambda_c = 1.6488(1)$,
there is a continuous phase transition from the vacuum 
to an active steady state. 
We introduce disorder by randomly removing a fraction $x$ of the sites.  
That is, for each $(i,j)$ $\in {\cal Z}^2$
there is an independent random variable $\eta (i,j) $ taking values
0 and 1 with probability $x$ and $1-x$, respectively.  The DCP is simply
the contact process restricted to sites with $\eta (i,j) = 1$; those having
$\eta (i,j) = 0$ are never occupied.  (Thus if exactly $m$ neighbors of
a given site have $\eta (i,j) = 1$, the creation 
rate at that site is at most $m \lambda /4$.)
Naturally, $1-x$ must exceed the percolation
threshold $p_c = 0.5927$ for there to be any possibility of an active state,
since on finite sets the CP is doomed to extinction.
 
Following the approach of Grassberger and de la Torre \cite{GRASS}, we study 
a large ensemble of trials, all starting from a
configuration very close to the absorbing state: 
a single particle at the origin. 
For $\lambda > \lambda_c (x)$ there 
is a nonzero probability that the process survives
as $t \rightarrow \infty$; for $\lambda \leq \lambda_c (x)$ 
the process dies with probability 1.
Of prime interest are $P(t)$, the survival 
probability at time $t$, $n(t)$, the mean number
of particles (averaged over all trials, including 
those that die before time $t$), and  $R^2(t)$, the 
mean-square distance of particles from the origin. 
At the critical point of the pure CP, these quantities 
follow asymptotic power-laws,    
\begin{eqnarray}
P(t) \propto& t^{-\delta} \\
 n(t) \propto& t^\eta \\
 R^2(t)\propto& t^z.
\label{eq:abpnr}
\end{eqnarray}
 
\noindent The exponents satisfy the hyperscaling relation
\begin{equation}
4 \delta + 2 \eta = d z,
\label{hyper}
\end{equation}
in $d \leq 4$ dimensions \cite{GRASS}.
For $\lambda < \lambda_c$, $P(t)$ and $n(t)$ decay exponentially, while for 
$\lambda > \lambda_c $,
\begin{equation}
P(t) \rightarrow P_{\infty} \sim (\lambda - \lambda_c)^{\beta'},
\label{betaprime}
\end{equation}
and $n(t) \sim t^d$.  
Thus log-log plots of $P(t), n(t)$ and $R^2(t)$ approach
straight lines at the critical point, and show a positive
or negative curvature in the supercritical or subcritical 
regimes, respectively.  
We adopt the  working hypothesis that the critical point of the DCP is also
characterized by asymptotic power laws.

We studied dilutions $x=$ 0.02, 0.05, 0.1, 0.2, 0.3, and 0.35.  Samples of from
$10^5 $ to $2 \times 10^6$ trials were generated for each 
$\lambda$ value of interest, each trial extending to a maximum 
time $t_{max} \leq 7 \times 10^4$.  
(As is usual in this sort of simulation, the time increment associated with an 
elementary event --- creation or annihilation --- 
is $\Delta t = 1/N$, where $N$ is
the number of particles.)
An independent realization of
disorder (the variables $\eta(i,j)$), is generated for each trial.
Our results for $P(t)$, $n(t)$ and $R^2(t)$ (see Fig. 1), allow us to
compute the local slopes $\delta (t)$, $\eta (t)$, 
and $z(t)$, using least-squares fits
to the data (in a logarithmic plot), distributed symmetrically about a given
$t$ (typically in the interval [$t/3$, $3t$]). We estimate the exponents by 
plotting the local slopes versus $1/t$
and extrapolating to $1/t \rightarrow 0$ \cite{GRASS89}, as in Fig. 4.
The critical point $\lambda_c(x)$ is determined by the criteria 
that at long times, 
log-log plots of $P(t)$ appear straight,
and that local-slope plots are free of marked curvature.  
Our results for the critical point are given in Table I.  
For small $x$, $\lambda_c(x) \approx \lambda_c(0)/(1-x)$,
as predicted by mean-field theory \cite{MARQUES,DCPST}.

Perhaps the most surprising feature of the critical DCP 
is the population size $n(t)$.
In the (pure) critical CP $n(t)$ 
increases monotonically with time; $\eta$ in Eq. (2) is positive.
In the critical DCP $n(t)$ decays at long times; $\eta$ is {\em negative}.  
The critical point evolution of $P$, $n$, and $R^2$, shown for several dilutions
in Fig. 1, has a regime characterized by pure CP scaling, followed by
a long crossover period leading to a different set of power laws.  The
apparently disparate results obtained for different dilutions are 
in fact described 
by a single function. 
For $n(t)$, this can be seen by shifting the data (plotted on log-log scales), 
to superimpose the maxima, which causes all of the points to fall 
close to a single
curve (see Fig. 2).  If $\tau(x)$ denotes the time at which $n(t;x)$
attains its maximum, $n_{max}(x)$, then we have

\begin{equation}
n(t;x) = n_{max}(x) {\cal N}(\tilde{t}) \;,
\label{nsc}
\end{equation}

\noindent where $\tilde{t} = t/\tau $ and ${\cal N}$ is a scaling function.  
We regard $\tau$ as the {\em crossover time} separating pure CP scaling
(for small $\tilde{t} $) from disordered CP critical behavior
($\tilde{t} \rightarrow  \infty $).  To scale the data for $P(t)$ and
$R^2(t)$ we use the same $\tau(x)$ values as for $n$, and fix
dilution-dependent amplitudes $A_P (x)$ and $A_{R^2} (x)$ by optimizing the data
collapse.  Thus for the survival probability

\begin{equation}
P(t;x) = A_P(x) {\cal P}(\tilde{t}) \;,
\label{psc}
\end{equation}

\noindent and for the mean-square spread,
\begin{equation}
R^2(t;x) = A_{R^2}(x) {\cal R}(\tilde{t}) \;.
\label{r2sc}
\end{equation}

To test these scaling expressions we plot $\tilde{P} \equiv P(t;x)/A_P(x)$,
$\tilde{n} \equiv n(t;x)/n_{max}(x) $ and 
$\tilde{R}^2 \equiv R^2(t;x)/A_{R^2}(x)$
versus $\tilde{t} $ in
Fig. 2.  Data for different dilutions collapse nicely onto a single curve.  
In Fig. 3 we plot the crossover time $\tau$
and the amplitudes versus $x$.  For $x\geq 0.1$ these 
parameters are well-described by power laws, for example, 
$\tau \propto x^{-3.2(1)} $.  (Figures in parentheses denote uncertainties.  
The corresponding exponents for $A_P$, $n_{max}$,
and $A_{R^2}$ are 1.67(4), -0.56(2) and -3.64(3), respectively.)  
Given the limited range of $x$, there is no reason to attribute any fundamental 
significance to these power laws.  The rapid decrease in $\tau$
with increasing $x$  can be understood by noting that
for weak disorder, the process must spread over a certain 
area before randomness becomes manifest; 
for small $x$, sizable regions of the lattice look nearly regular.
Thus for $x \leq 0.2$, $n(t)$ and $R^2(t)$
have not reached their asymptotic behavior 
at the end of the trials ($t_{max} = 7 \times 10^4$).
Since $n(t;x=0.02)$ does not exhibit a maximum on 
simulation time scales, the data for $x=0.02$ are not included in the
scaling analysis.

Given that the approach to asymptotic scaling 
is more rapid for larger dilutions,
we turn first to the data for $x=0.35$ to estimate 
critical exponents.  Fig. 4, for $\eta (t)$, shows that the
local slope undergoes a marked change during the evolution. 
($z(t)$ also changes over a very wide range.)
At long times the local slopes for $x=0.3$ and $x=0.35$ exhibit steady trends, 
permitting reliable extrapolation 
of their asymptotic values; for smaller dilutions  
only $\delta$ can be estimated with confidence (see Table I). 
The scaling plots of Fig. 2 show that the scaling functions 
${\cal P}$, ${\cal N}$, and ${\cal R}$ follow power laws with pure-DP
exponents for small $\tilde{t} \equiv t/\tau$, and that for large $\tilde{t}$
they approach power laws with the exponents 
found in the local-slope analyses.

We also determined the 
ultimate survival probability exponent $\beta'$ (defined 
in Eq. (\ref{betaprime})), for dilutions 0.1 and 0.2.  
While $\beta'$ is conceptually distinct from the order-parameter 
exponent $\beta$, defined through $\rho \sim (\lambda - \lambda_c)^{\beta} $,
these exponents are equal in the CP and allied models.
Stationary-state simulations support the relation $\beta = \beta'$ 
for the DCP as well \cite{DCPST}.

Table I reveals the dramatic effect 
dilution has on the exponents. 
Comparing the critical exponents for 
the DCP with those of the pure model, it is 
evident that the shifts all reflect a diminished 
ability of the critical process to survive and 
expand in the presence of disorder.  While this accords with 
intuition, the small $z$ value, and negative $\eta$
are unexpected.  Most absorbing-state
transitions have $\eta \geq 0$, but this does not appear to be a fast
rule, so long as $\eta_s \equiv \eta + \delta > 0$. 
($\eta_s$ governs the population in {\em surviving} trials.)   
Negative values of $\eta$ were also found in a 
(disorder-free) model with multiple  
absorbing configurations \cite{SNR}.  
Equally puzzling is the massive violation of hyperscaling, Eq. (\ref{hyper}).
In fact, hyperscaling in the DCP may be
{\em incompatible} with the simple requirement $\eta_s > 0$; for $d=2$,
Eq. (\ref{hyper}) implies $\eta_s = z - \delta$, 
which in turn yields (for $x=0.35$),
$\eta_s \simeq -0.35$.  Alternatively, if
we take $\eta = -\delta$, (that is, as small as possible without
forcing $\eta_s < 0$), Eq (\ref{hyper})
yields $z = \delta$, in marked disagreement with the data.  

One might regard violation of hyperscaling
as evidence that our simulations have yet 
to probe the asymptotic scaling regime.
While simulations which run to a finite time cannot rule this out, 
the data suggest no trend 
that would bring the exponents into agreement with hyperscaling.  
Another interpretation
\cite{DANI} is that sample-to-sample fluctuations and 
lack of self-averaging lead to a violation 
of the scaling hypothesis itself (as set out, e.g., 
in Ref.\cite{GRASS}).  We hope to
resolve this issue in a future study.

Bramson, Durrett and Schonmann studied a one-dimensional CP with disorder
in the form of a death rate randomly taking one of two values (independently)
at each site \cite{BRAMSON}.  They demonstrated 
that this model possesses an intermediate phase in which
survival (starting, e.g., from a single particle) 
is possible, but the active region grows
more slowly than linearly. 
In two or more dimensions, Bramson et al. 
conjectured, there is no intermediate phase.
Our results support this conjecture.  Simulations
at $x=0.1$, with $\lambda$ close to, but slightly above $\lambda_c$
(to be precise, $\lambda = 1.86$ and 1.87, corresponding 
to $(\lambda -\lambda_c)/\lambda_c
= 0.008$ and 0.014, respectively),  showed  
$n(t) \sim t^2$ (and similarly for $R^2(t)$),
consistent with the radius of the active region growing $\sim t$.  
Thus a sublinear growth phase, if it exists at all, is confined
to a very narrow range of creation rates.  While our model
incorporates dilution rather than a random death rate, one would expect such an
intermediate phase to be a rather general feature of disordered 
contact processes, so that its apparent absence here argues for the validity
of the conjecture.

\begin{table}
\caption{\sf Critical parameters and exponents obtained through simulations of
 the DCP. The numbers in brackets indicate uncertainties.}
\begin{center}
\begin{tabular}{|c|c|c|c|c|c|} 
$x$     & $\lambda_c$ & $\beta'$  & $\delta$ & $\eta$    & $z$   \\ \hline\hline
$0$      & 1.6488[1]     & 0.586[14]& 0.460[6] & 0.214[8] & 1.134[4] \\
$0.02$ & 1.6850[3]     & 0.566[7]  & 0.467[1] & 0.216[3] & 1.104[2] \\
$0.05$ & 1.7405[3]     &                & 0.59[1] &            &              \\
$0.1$   & 1.8448[4]     & 0.70[1]     & 0.59[1] &              &              \\
$0.2$   &  2.097[2]      & 0.72[2]      & 0.60[2] &            &              \\
$0.3$   &  2.437[2]      &               & 0.59[1]  &  -0.43[3] & 0.27[3]  \\ 
$0.35$  & 2.655[2]      &                & 0.59[1] &  -0.41[3] & 0.24[3]   \\ 
\end{tabular}
\end{center}
\label{results}
\end{table}
 
In summary, we find that quenched disorder induces a radical change in 
the critical exponents describing spreading in the contact process.  As the 
population spreads in the critical process,
there is a crossover from pure CP behavior to a new set of critical exponents.
The crossover time appears to diverge as the dilution tends to zero.  A
related finding is the demonstration by Noest of
anomalously slow dynamics, in the form of power-law relaxation to the vacuum,
in disordered directed percolation \cite{NOEST88}.
Although our results on scaling are restricted to dilutions 
$0.05 \leq x \leq 0.35$,
we expect that such a crossover occurs for all $0 < x < 1 - p_c$,
albeit at very long times for small $x$.
One is inclined to suppose that the DCP 
is but one member of a universality class
encompassing all disordered models with a continuous transition to a
unique absorbing configuration.  In support of this we observe that
$\delta$ (the only exponent
we are able to determine over a wide range of $x$ values), is
insensitive to the degree of dilution, and that evolutions for
various $x$, with very different crossover times, are described by
common scaling functions.  Studies of absorbing state transitions in other 
disordered models are needed to verify the universality hypothesis.
Finally, we note that our result, $\beta = \beta'  = 0.71(2)$, is
incompatible with that reported by Noest ($\beta = 1.10(5)$)
for disordered DP in 2+1 dimensions \cite{NOEST}.
We shall return to this question in a future work on the static behavior of the
diluted contact process.
 
\begin{flushleft}
{\bf Acknowledgments}
\end{flushleft}
We thank Dani ben-Avraham for extensive discussions and Geoff Grinstein for
helpful suggestions.
A.G.M. was supported by CNPq (Brazil). She thanks the staff of Lehman
College, City University of New York, for their hospitality during her visit.
\vspace{2em}

\noindent $^a${\small e-mail address: dri@fisica.ufmg.br }\\
$^b${\small e-mail address: dickman@lcvax.lehman.cuny.edu }

\newpage

\noindent{\bf Figure Captions}
\vspace{1em}

\noindent FIG. 1.  Survival probability $P$, mean 
population $n$, and mean-square
distance of particles from the origin $R^2$, in the 
critical contact process with dilution $x$.
$\diamond$: $x=0.05$; $+$: $x=0.2$; $\times$: $x=0.35$.
\vspace{1em}

\noindent FIG. 2. Scaling plots of spreading 
in the DCP.  Upper panel: $\ln \tilde{P} $
vs $\ln \tilde{t}$; the straight lines have 
slopes of -0.46 and -0.60.  Middle: $\ln \tilde{n} $;
the slopes are 0.241 and -0.42.  Bottom: 
$\ln \tilde{R^2} $; the slopes are 1.134 and 
0.24.  Symbols: $\diamond $: $x = 0.05$; $\Box$: $x=0.10$; $+$: $x=0.20$; 
$\triangle$: $x=0.30$; $\times$: $x=0.35$.
\vspace{1em}

\noindent FIG. 3. Log-log plots of $\tau$ 
($\diamond$), $n_{max}$ ($\Box$), $A_P$ ($+$),
and $A_{R^2}$ ($\times$) vs dilution $x$.
\vspace{1em}

\noindent FIG. 4. Local slope $\eta(t)$ 
versus $t^{-1}$ in the diluted contact process with $x=0.35$. Symbols: $+$: $\lambda = 2.650$; $\times$: $\lambda = 2.655$; $\diamond$:
$\lambda = 2.660$.
\vspace{1em}

\end{document}